# PEvoLM: Protein Sequence Evolutionary Information Language Model

Issar Arab
Department of Computer Science, University of Antwerp, Antwerp, Belgium
Biomedical Informatics Network Antwerpen (biomina), 2020 Antwerp, Belgium
Email: issar.arab@tum.de

*Abstract*— With the exponential increase of the protein sequence databases over time, multiple-sequence alignment (MSA) methods, like PSI-BLAST, perform exhaustive and time-consuming database search to retrieve evolutionary information. The resulting position-specific scoring matrices (PSSMs) of such search engines represent a crucial input to many machine learning (ML) models in the field of bioinformatics and computational biology. A protein sequence is a collection of contiguous tokens or characters called amino acids (AAs). The analogy to natural language allowed us to exploit the recent advancements in the field of Natural Language Processing (NLP) and therefore transfer NLP state-of-the-art algorithms to bioinformatics. This research presents an Embedding Language Model (ELMo), converting a protein sequence to a numerical vector representation. While the original ELMo trained a 2-layer bidirectional Long Short-Term Memory (LSTMs) network following a two-path architecture, one for the forward and the second for the backward pass, by merging the idea of PSSMs with the concept of transfer-learning, this work introduces a novel bidirectional language model (bi-LM) with four times less free parameters and using rather a single path for both passes. The model was trained not only on predicting the next AA but also on the probability distribution of the next AA derived from similar, yet different sequences as summarized in a PSSM, simultaneously for multi-task learning, hence learning evolutionary information of protein sequences as well. The network architecture and the pre-trained model are made available as open source under the permissive MIT license on GitHub at https://github.com/issararab/PEvoLM.

## I. Introduction

In contemporary computational biology and, more specifically, in the area of proteins and the prediction of their properties, sequence alignment forms the de-facto standard input to nearly all machine learning methods [1]. Multiple sequence alignment methods, or MSAMs, are a set of algorithmic solutions for the alignment of evolutionarily related sequences. They can be applied to DNA, RNA, or protein sequences. Those algorithms are designed to take into account evolutionary events such as mutations, insertions, deletions, and rearrangements under certain conditions [2]. The main functionality of these alignment techniques is to search for homologs of a query sequence in a database of protein sequences, as they tend to share structure and function. For the past two decades, training machine learning models with evolutionary information representations, generated by multiple sequence alignments, has revolutionized the prediction power of AI methods. Multiple aspects of protein function and structure were studied and investigated following the same approach and achieved significant results in the prediction performance. Such downstream-specific tasks include protein secondary structure [3, 4, 5, & 6], transmembrane protein regions [7, 8, & 9], inter-residue contacts [10], and sub-cellular localization predictions [11, 12] as well as protein to protein interactions [13, 14, & 15]. However, this increase in performance has become costly in recent years, with the continuous exponential growth of bio-sequence data pools. UniProt is one example of such data stores, in which the entries keep doubling every couple of years [16].

The rise of machine learning and deep learning has undeniably engendered a paradigm shift, revolutionizing and bestowing remarkable advancements in various domains of bioinformatics, such as mass spectrometry [17], protein sequences and evolutionary information [18], and cardiotoxicity liability predictions [19, 20]. To cope with such tremendous growth of bio-sequence repositories, alternative approaches using artificial intelligence are then researched among the community. One prominent solution that can compete with conventional search methods is the direction of Embedding Language Models (ELMo) [21], a state-of-the-art technology borrowed from the NLP field.

In the NLP setting, pre-trained word representations are a central component to several natural language comprehension models [22, 23]. However, learning high-quality representations is a difficult task. Ideally, these representations have to model both the dynamic features of word usage, like semantics and morphology, and how they differ across linguistic domains, like polysemy modeling. Pre-trained word vectors [22, 23, & 24] learned from a large corpus of unlabelled content have the ability to model these syntactic and semantic word representations. They represent the core of many state-of-the-art NLP architectures out there, such as semantic role labeling [25], question answering [26], and textual entailment [27].

An ELMo is trained on a large corpus of unlabelled natural text, Wikipedia as an example [2], to predict the next most probable word in a sentence given all the previously seen tokens. However, in a bi-directional language model, during training we learn the probability distribution of the next word in the sentence from both directions, i.e. predicting a pivot word given all the previous tokens from a forward pass and from a backward pass of a sentence. This bi-directional autoregressive [21, 28] paradigm has revolutionized NLP allowing the model to develop a syntactic and a semantic self-learning of the word in a sentence, a.k.a. the context. This means that, for a particular word, the model will provide different contextualized embeddings, depending on the sentence it is used in.

Given the close nature of protein sequences to natural language sentences, the same approach was adopted to train

SeqVec [29] on UniRef50, a corpus of 9.5 billion amino acids, which is around 10 folds larger than Wikipedia in terms of tokens(words). In their research work, Heinzinger et al. [29] proposed a novel embedding tool of protein sequences that replaces the explicit search for evolutionary-related proteins in a database. The model was trained on predicting the next amino acid in the sequence. The new approach can be described as an implicit transfer of biophysical and biochemical information learned during the training of a bi-directional language model embedder [21], inspired from NLP, on a large unlabelled set of sequences.

The predictive power of the embeddings was then tested on downstream tasks categorized under two levels: per-residue and per-protein predictions. The results showed that the models were able to reach a good performance, but did not outperform the state-of-the-art MSA-based tools. This paper's idea was then to train a novel bi-language model on PSI-BLAST's output with transfer learning, which would eventually encode evolutionary information of the proteins within its embedding representations, with the goal of boosting the final embedding power and potentially reduce the size of residue embedding while maintaining the amount of information encoded.

To train the new embedder, a large curated dataset of sequences was compiled with their corresponding PSSMs of size 1.83 Million proteins (~0.8 billion amino acids). The dataset of proteins is reduced to 40% sequence identity, with respect to the validation/test sets, and contains sequences ranging between 18 and 9858 residues in length. The next sections will include the research question, data and methods, followed by the experiments, results and discussion section.

All results presented in this paper were conducted on a remote Linux VM granted by Google with a system memory size (RAM) of 120GB and 32 Intel® Xeon® CPUs with a maximum speed of 2.3 GHz. The machine also contained a cluster of 8 Tesla V100 SXM2 GPUs with a dedicated memory of 16GB each, from which 2 GPUs were used to train the novel embedding language model.

## II. RESEARCH QUESTION

In probabilistic machine learning, a probabilistic model is a joint distribution of hidden variables, referred to as $z$, and observed variables, referred to as $x$. This statement can be written in probabilistic notations as $p(x, z)$.

In this setting, inference about the unknowns $z$ is done through what we call the *posterior* distribution. It is a conditional distribution of the hidden variables given the observations. This statement is translated in probabilistic notations as $p(z \mid x)$.

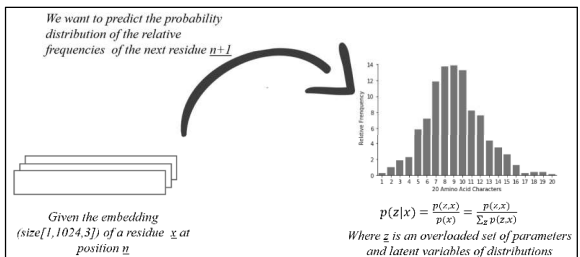

**Figure 1.** Problem statement visualization of the research goal on learning representations of protein evolutionary information

The goal of this research work is to go directly from the amino acid sequence of a given protein to a Position-Specific Scoring Matrix, more specifically to the matrix of relative frequencies. In other words, I want to translate a vector of characters of length $L$, here the protein sequence, to a matrix of dimensions $L$ x 20, which is the Position-Specific Scoring Matrix in this context. In this PSSM map, each row represents the relative frequencies of a residue in the sequence with respect to all the 20 known AAs. These row values represent a discrete probability distribution, where the statistics in each row sum up to 1 or 100 depending on the scale. My approach aims at learning those representations directly from the sequence following the same paradigm applied in autoregressive models [30]: From the observations of previous time steps given as an input, I want to make predictions of the next time step.

Figure 1 visualizes the main problem statement in this research work, which aims to learn protein evolutionary information. In a protein sequence of length $L$, given the embedding of a residue $x$ at position $n$ as well as the previous residues information passed on, I want to predict the probability distribution of the relative frequencies of the next residue $n+1$. In mathematical notations, this is equivalent to a conditional probability distribution $p(z \mid x)$, where $z$ is an overloaded set of parameters and latent variables of the distribution. Since the evolutionary information embedded within the relative frequencies matrix at the residue level is a discrete probability distribution and the evidence can be written as the marginalization of the joint distribution over $z$, the probabilistic model can be reformulated as follows:

$$p(z|x) = \frac{p(z,x)}{\sum_z p(z,x)}$$

To solve this model and since the evidence is not tractable, I appeal to approximating my posterior inference through Variational Inference (VI) [31] using Kullback-Leibler divergence ($\mathbb{KL}$-Div). The main idea is to: (1) find a tractable distribution $q(z|v)$ that is similar to $p(z|x)$, and then (2) use $q(z|v)$ to answer the questions about $p(z|x)$ that I care about. Here, I just have to find the optimal parameters $v^*$ that minimize the $\mathbb{KL}$-divergence.

Following the universal approximation theorem [32], I opted for a deep neural network as a parametric complex function that can learn any distribution given enough neurons in a two-layer network. Therefore, I converted the inference to an optimization problem. The final loss function that I need to minimize is:

$$v^* = \arg\min_v [\sum p(z|x)\log(p(z|x)) - \sum p(z|x)\log(q(z|v))]$$

## III. DATA AND METHODS

Running PSI-BLAST on a large set of protein sequences from UniProt [33], for example Reference Cluster with 50% sequence identity, was unrealistic as the search is highly exhaustive and might take months, depending on the available computing resources, for a couple of millions of protein sequences. Therefore, I opted for collecting the cached PSSMs from Predict-Protein [34], an Internet service for sequence analysis as well as prediction of protein structure and function. Figure 2 depicts the high-level integration and transformation steps applied in the pipeline used for data gathering and pre-processing. Protein sequences from UniProt [33] Reference Cluster with 50% sequence identity (uniref50 2019_12), ~38.8 Million proteins, were checked one by one, and for the matched hits I retrieved PSSMs with evolutionary information along with their corresponding alignment files. All data was cleaned and pre-processed for high-quality PSSMs to use during training. The final set of 1.83 Million protein sequences (~0.8 billion amino acids) and their corresponding PSSMs, along with validation, and test sets have been made made public to the community and deposited to Zenodo at (https://zenodo.org/record/4300971).

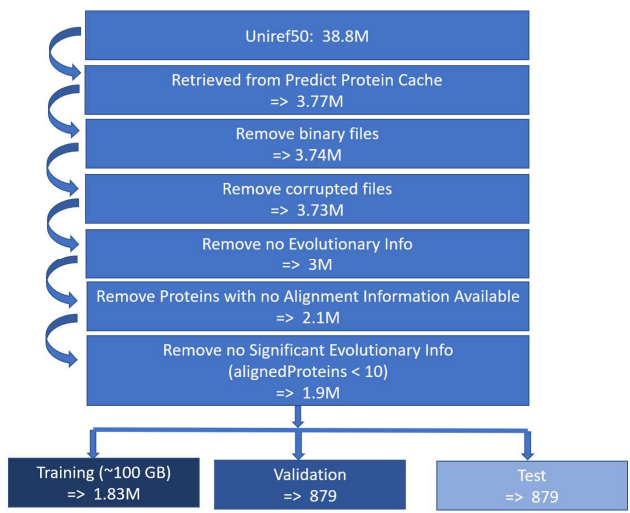

**Figure 2.** Simplified visualization of the gathering and pre-processing pipeline of the whole Uniref50 dataset to obtain the final training validation and test sets

To analyze my development set (~2.1M sequences) in terms of the amount of evolutionary information encoded in the relative frequencies' matrices, I examined the distribution of protein sequences with respect to the family size of the number of aligned proteins for each sequence in the set. The analysis showed that the majority of protein PSSM files are built on family sizes of more than 1000 aligned sequences while ~200k PSSMs are built on families of size less than 10 aligned sequences.

To maintain a high-quality training set, I conducted further analysis to set a threshold of the family size for which proteins will be either discarded or used in the final training. To pick that threshold, I calculated the complement of the cumulative distribution at different threshold values (i.e. family sizes). Mathematically speaking, I computed the distribution of 1- F(x), where F(x) is the CDF of the number of aligned proteins to each sequence in the development set. The analysis showed that 9% of the sequences have less than 10 aligned proteins, leaving us with 91% of the total development set with significant encoded evolutionary information. Therefore, the threshold of 10 was picked to discard such sequences with no significant evolutionary information.

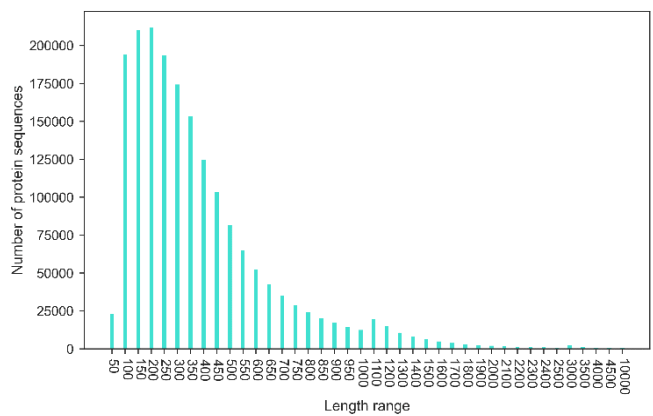

**Figure 3.** Proteins sequence length distribution in the final training set (~1.83 M protein sequences)

In ML-based NLP algorithms, two major pieces of information are helpful in the model architecture design and the hyperparameters selection, especially when it comes to the size of mini-batch training enforced by the GPU memory constraint. The first represents the distribution of proteins in my final training set by their sequence length (Figure 3). The histogram displays the length distribution I am dealing with in the training set, and it characterizes a right-skewed distribution with a mean of around 300 residues. The second important piece of information is the distribution of vocabulary in my training set. Figure 4 shows the amino acids composition of my dataset, which is a rather right-skewed distribution than a uniform one. This information will be used to compute the cross-entropy (CE) random baseline for predicting the next amino acid in the sequence.

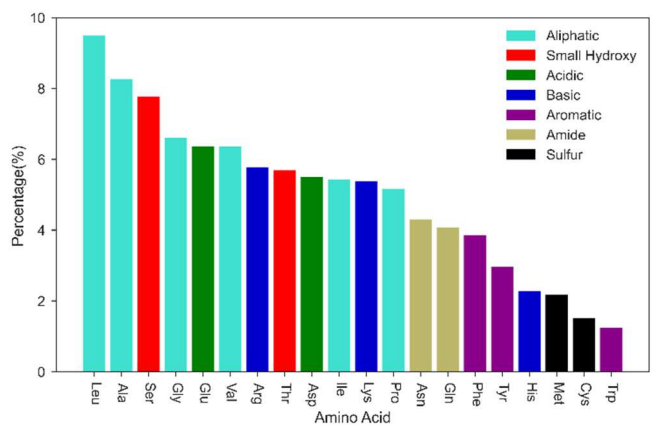

**Figure 4.** Distribution of the amino acid composition in the final training set (~0.8 billion amino acids)

## IV. PEvoLM Architecture

PEvoLM core technology is LSTM, which is the same machinery used in the initial bi-directional autoregressive approach presented by Peters et al. [21]. ELMos are known to require both large datasets and significant training time and resources to reach convergence. Therefore, I opted for transfer learning using SeqVec/ELMo [29], a sequence embedder trained on a corpus of 9.5 billion amino acids to predict the next residue in a sequence. The architecture provides a per-residue embedding of size (3 x 1024).

Initial experiments showed that a simple bi-directional LSTM succeeds in perfectly learning a mapping function on very small training sets, i.e. a couple of hundreds, but struggles to improve the performance on a large dataset. To boost the capacity of a network, one can either go deeper by increasing the number of layers, go wider by increasing the number of neurons, or both. A joint research work between a faculty member at the University of Toronto and a scientist at Microsoft Research [35] has shown empirically that shallow feed-forward networks can learn the complex functions formerly learned by deep neural networks and obtain performances that were previously only possible with deep architectures. This conclusion was convincing enough to drop the idea of going deeper, in terms of layers, but rather look towards the direction of increasing the number of hidden units.

An LSTM architecture contains several components named *memory blocks*. Such blocks are called gates controlling the information flow in a network, including the input, output and forget gates. The LSTM structure is uniquely defined by the number of its input and output units. To raise the capacity of an LSTM, one can just increase the hidden size units, hence increasing the capacity of the cell state to carry more information along to the next time steps for complex tasks. However, those numbers determine the computational complexity for training an LSTM network. For a moderate number of hidden input units, dimensionality, i.e. the complexity of the network, is largely dominated by the size of the output hidden units. Sak et al. [36] from Google labs suggested an alternative approach that addresses the complexity of large-capacity LSTMs. They proposed the addition of a linear projection layer applied right after the output and before the recurrent connection to the cell. Therefore, the recurrence is applied from a smaller projected hidden state, which ultimately reduces the number of computations within the whole LSTM cell, while maintaining the large capacity of the cell state. The later described module is referred to as LSTM with recurrent projection layer (LSTMP) [36]. This technology has been adopted by Jozefowicz et al. [37] in Google Brain labs to train different variations of language models on a very large corpus of 0.8 billion words and a vocabulary size of 793471 words [38]. Jozefowicz et al. [37] empirically proved that when trying to fit an LSTM network architecture on very large and complex datasets, the size of the LSTM heavily matters.

To improve the LM architecture, I customized the CUDA-optimized Long-Short Term Memory cell implemented with TorchScript [1] to include a recurrent linear projection layer. For the hidden and projection sizes, I decided to go with half the dimensions used in [21 & 39], i.e. 2048 units and 256-dimension projections.

Inspired by the original ELMo paper by Peters et al. [21], a residual skip connection was added from the first to the second LSTM layer, with the goal of boosting the training performance. To further increase the capacity of the network and have more control over the memory size allocated by the LSTM layers for variable input sizes, I also added a fixed size (1024 hidden unit) non-linear input layer, to the LSTM cell, with LeakyReLU as an activation function. Concerning the weights initialization of the LSTM matrices, I initialized the input projection layer with a Kaiming [40] uniform initialization, the biases of the LSTM gates to a value of 1.0, as it was shown to perform well for long size dependencies [39], and the rest of the matrices were initialized with a uniform distribution of a standard deviation $\sigma = \frac{6.0}{hidden\_size + projection\_size}$.

Combining all the LM state-of-the-art technologies discussed in this subsection, a complex network architecture was designed as visualized in figure 5. The model comprises 2 layers of LSTMs, with projections, stacked one after the other. The first layer takes as input SeqVec uncontextualized embeddings concatenated with a one-hot encoding vector of size 20, making it a total input size of 512+20 = 532 units. The second LSTM layer takes as input contextual representations concatenating the output of the first layer with the 2-layer context-aware embeddings from SeqVec, resulting in an input size of 512*2+256 = 1280 units. As for the residual block, the two representations of size 256 from both layers are then summed elementwise to serve as input to the two parallel linear layers, one for predicting the next amino acid and the second for inferring the next PSSM column distribution.

Concerning the loss functions used in the architecture, we have both a cross-entropy loss assessing the predictions of the next amino acid and a $\mathbb{KL}$-Divergence loss measuring the closeness of the predicted discrete probability distribution to the ground truth relative frequencies of the next PSSM column. This corresponds to an optimization problem with respect to multiple objective functions. The commonly adopted approach to solve such multi-task problems computes a weighted linear sum [41] combining all the objective losses:

$$\begin{cases} L_{total} = \sum_i \alpha_i L_i \\ Where \sum_i \alpha_i = 1 \end{cases}$$

Here, $L_i$ is the objective loss of task $i$ and $\alpha_i$ is the corresponding weight given to that loss. I want to point out here that my primary goal is to learn evolutionary information representations. Additionally, previous experiments showed that the CE loss was more dominant in the final objective function when assigning a similar weight, being nearly 4 times larger than the $\mathbb{KL}$-divergence loss. Therefore, I decided to bring both losses to the same scale, which will also give more weight to the $\mathbb{KL}$-divergence loss. Tuning the loss coefficients, the final objective function is defined as:

$$L_{final} = 0.25 * CE_{loss} + 0.75 * KL_{loss}$$

---

[1] https://pytorch.org/blog/optimizing-cuda-rnn-with-torchscript/

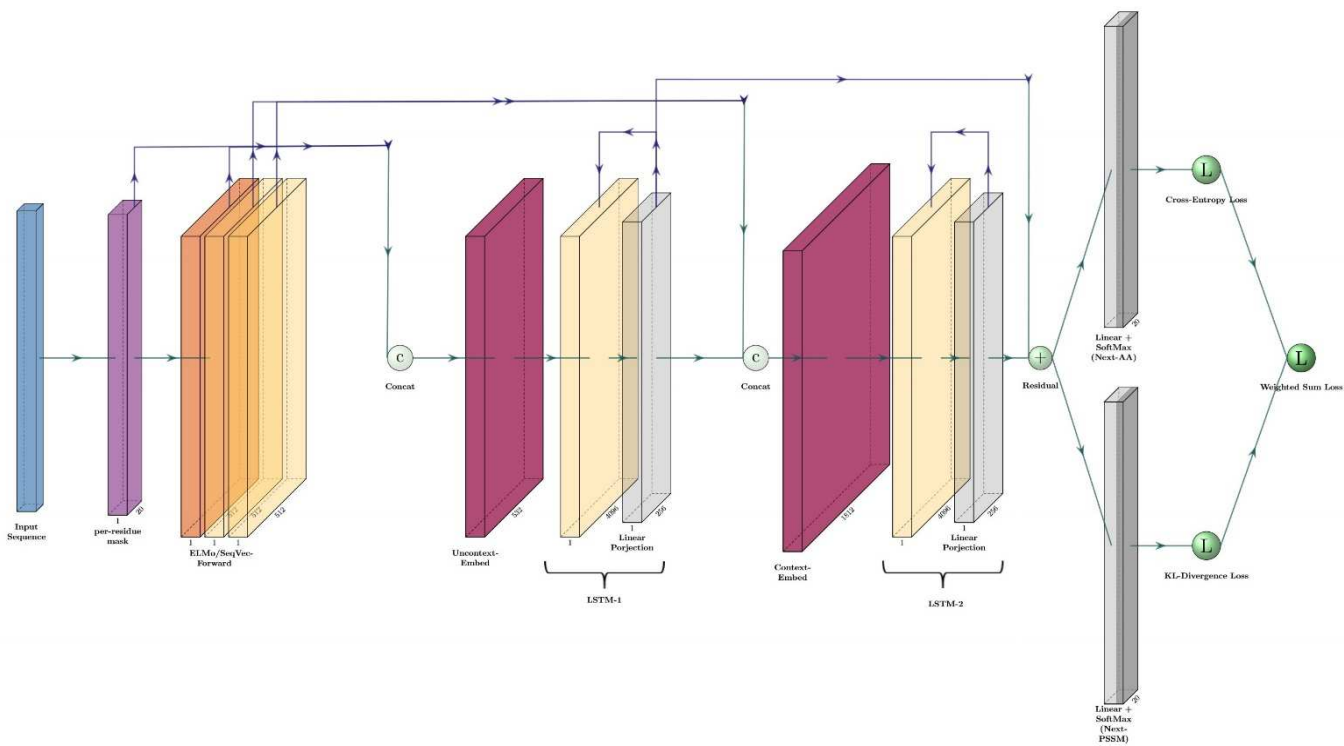

**Figure 5.** Conceptual visualization [2] of the novel single path 2-layer bidirectional embedding language model architecture. The model is trained on predicting both the distribution of the next amino acid and its corresponding evolutionary information relative frequencies. The architecture makes use of LSTMs with projection and a residual connection

## V. RESULTS AND DISCUSSION

Hyperparameter tuning on the presented architecture revealed that the best-performing model is trained with truncated backpropagation through time of 100 timesteps and a batch size of 128 sequences.

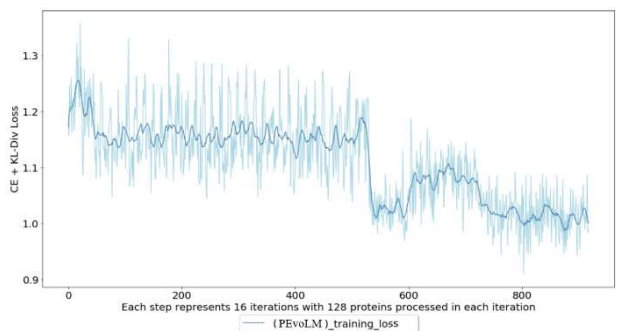

**Figure 6.** Smoothed loss training curve of PEvoLM multi-task learning on a dataset of 1.83 million proteins using a maximum time step of 100 residues.

The model training took 1 week, and the converged training curve for the joint final loss is displayed in Figure 6. The curve shows how the high-capacity architecture has allowed the model to learn representations from the 1.83 million sequences. From the training curve, I clearly see that the loss function has made 3 improvement drops: The first around 800 iterations after processing 100K sequences, the second around 8300 iterations after processing 1 million sequences, and the third 12K iterations after processing 1.5 million proteins. To assess how well each of the separate tasks, predicting either the next residue or the next PSSM column did contribute to the final loss, I plot the training curves of each task separately and compare them to baselines.

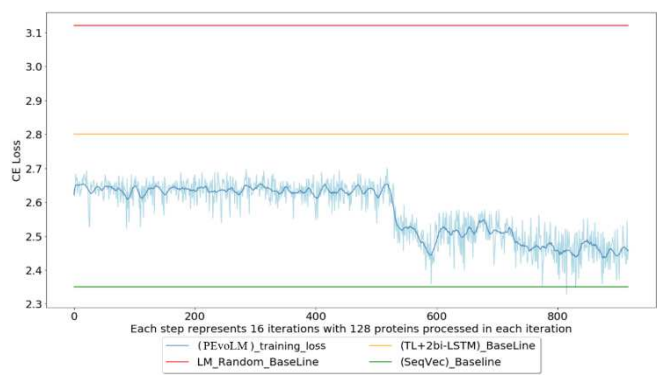

**Figure 7.** PEvoLM smoothed loss training curve predicting the next amino acid in the sequence. The model was trained on a dataset of 1.83 Million proteins using a maximum time step of 100 residues. The training curve is compared with three baselines.

Figure 7 shows the training curve (blue) of the final self-learning task predicting the next amino acid in a sequence from all previously seen residues. Comparing the learning trend with the one using plain LSTMs (Orange baseline), we see how the large capacity architecture did improve the training significantly when compared to plain architectures. While the initial plain LSTM architecture's training maintained a constant rate (~2.8) over time, the training loss of the final architecture starts at a much lower value (~2.65), with a low range oscillation throughout the whole training and converged at a loss of ~2.4, which is close to SeqVec reported performance of ~2.35 (green baseline).

---

[2] https://github.com/issararab/PEvoLM/blob/master/img/LM_architecture.jpg

throughout this research, achieving a joint best loss of 1.0, with a cross-entropy converging around CE = 2.4 and a $\mathbb{KL}$-Divergence loss value at $\mathbb{KL}$-Div = 0.5. The best model architecture is defined with a hidden size of 2048 and an output projection size of 256 units. The embedding predictive power was further evaluated on two categories of downstream tasks: the first task involves secondary structure, which is a per-residue type of predictions; the second comprises subcellular localization and soluble vs. membrane proteins as a per-protein level of predictions.

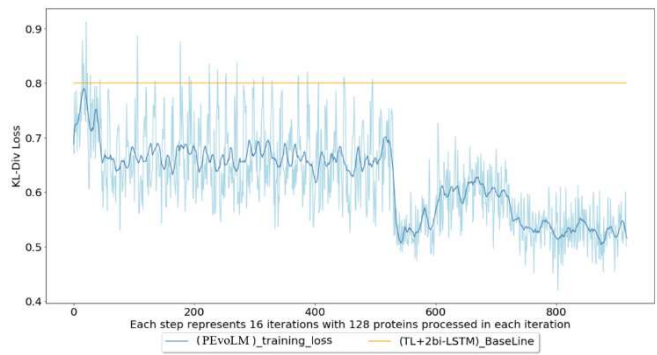

**Figure 8.** PEvoLM smoothed loss training curve predicting the next PSSM column in the sequence. The model was trained on a dataset of 1.83 million proteins with corresponding relative frequencies using a maximum time step of 100 residues. The training curve is compared with the vanilla LSTM architecture baseline.

Similarly to the cross-entropy loss, the $\mathbb{KL}$-divergence showed a comparable behavior for the novel ELMo architecture. Figure 8 displays the next PSSM column smoothed training curve where the model converged around 0.5, more than 1/3 lower than the plain LSTM architecture (orange). Table 1 summarizes the performance improvement

**Table 1.** The table compares the performance of the final ELMo architecture training 2 stacked bidirectional LSTMs with projection via transfer learning (TL). The table shows the distinct performance metrics of the CE and $\mathbb{KL}$-Divergence losses as well as the joint final loss. Besides, the table displays the values of the random, plain architecture, and SeqVec reported baselines.

|  | Cross-Entropy loss (AA) | $\mathbb{KL}$-Divergence loss (PSSM) | Joint loss (AA + PSSM) | Hidden size / Output size |
|---|---|---|---|---|
| Random baseline | 3.12 | - | - | - |
| SeqVec reported baseline | **2.35** | - | - | 4096/512 |
| TL+2bi-LSTM | 2.8 | 0.8 | 1.8 | 256/256 |
| **PEvoLM** | 2.4 | **0.5** | **1.0** | 2048/256 |

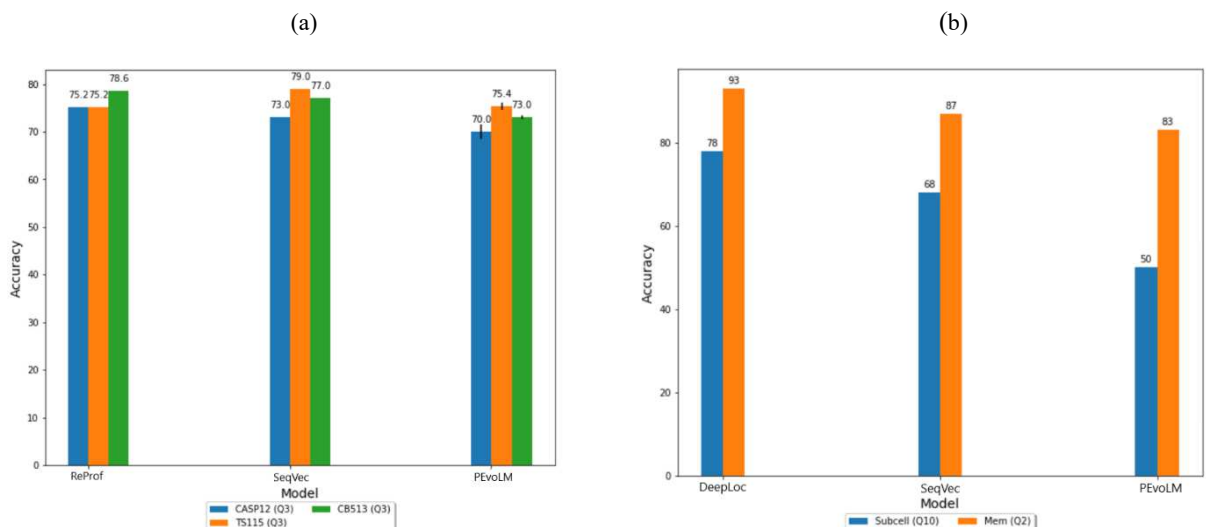

**Figure 9.** (a) 3-state secondary structure prediction comparison between MSA-based and ELMo-based inducers (b) Subcellular localization and membrane vs soluble protein prediction comparison between MSA-based and ELMo-based inducers

Regarding the first evaluation, a model was trained using the new embeddings as input to predict the three states of a protein secondary structure: helix (H), strand (E), and coil (C). For a fair benchmarking, the downstream task model architecture was similar for all three compared methods. The embeddings were evaluated on three test sets:

- $TS_{115}$ [42]: a set of 115 sequences derived from high-quality protein structures (i.e. < 3 Å) with no more than 30% PIDE to any protein of known structure in the PDB [43] in 2015.
- $CB_{513}$ [44]: a set of 513 non-redundant sequences compiled after a Structure Integration with

Function, Taxonomy and Sequence (SIFTS) [45] mapping.
- $CASP_{12}$ [46]: a set of 21 protein sequences retrieved in 2018 from the $CASP_{12}$ free-modelling targets after a SIFTS mapping.

Figure 9-a displays the performance results of PEvoLM embedder compared to SeqVec, another ELMo-based model, and ReProf [87], an MSA-based model. ReProf is built on PSSM matrices input generated by MSA methods. The latter is still considered as one of the state-of-the-art methods on this task. From the histogram plot, we see that SeqVec is performing quite well without the need of evolutionary information to make predictions. We also observe that SeqVec and PEvoLM are performing in a quite comparable way where SeqVec is still doing better. However, I should mention that SeqVec is relying on embeddings of size 3x1024 for each residue in the sequence to reach this performance, whereas PEvoLM requires embeddings with only half the size of its rival (3x512).

The second evaluation was conducted on predicting the membrane-bound proteins from the water-soluble ones. The two-state predictions were tested on a set of 846 proteins retrieved from DeepLoc [47] published supplementary data set. DeepLoc is a state-of-the-art tool relying on MSA profiles output to build its models. It also includes a model predicting the 10 states of subcellular localization, corresponding to the third benchmarking evaluation. The same annotated test set was used to evaluate and benchmark PEvoLM and SeqVec embeddings with DeepLoc (Figure 9-b). The orange bars correspond to the performance results of the $Q_2$ membrane predictions. We see that PEvoLM is lower by 10% and 4% compared to DeepLoc and SeqVec, respectively.

While DeepLoc is still outperforming both of the ELMo-based methods ( $Q_2$ =93%), the suggested embedding performance is still significantly high hitting an 83% with an embedding size of only 3x512 per residue. However, the $Q_{10}$ localization predictions seem to perform poorly, with an overall accuracy of 50%, which is 28% lower than DeepLoc and 18% lower than SeqVec. For full comparison of the performance of the presented LM and other experimented architectures, please refer to figure 4 & 7 presented by Elnaggar et al. [48].

## VI. CONCLUSION

The current algorithms for generating evolutionarily related information of protein sequences is largely dominated by multiple sequence alignment methods. We have seen that this technique is one of the most widely used modeling approaches in biology. MSAMs are used to expose those restricted evolutionary regions within a sequence. The results of these methods represent an essential input to several downstream applications in the field of bioinformatics. The process is simply described as searching for homologs of a query sequence in a database of protein sequences, capturing the conservation patterns in the alignment, and storing this information as a matrix of numerical scores for each position in the alignment.

Even though, the evolutionary information representations generated by MSA methods have revolutionized the prediction power of AI methods for the past two decades, this increase in performance has become costly in recent years, with the continuous exponential growth of bio-sequence data pools. Thanks to the similarity of protein sequences to natural language, NLP state-of-the-art algorithms, such as ELMo, has successfully been applied in bioinformatics.

The main objective of this research was to deploy an ELMo tool incorporating evolutionary information in its representations and offering a better trade-off between computing resources and runtime. While pre-training is costly, the inference is cheaper. A novel bidirectional language model was trained following the autoregressive approach. The model was trained on a set of 1.83 million proteins (~0.8 billion residues), predicting both the next amino acid in a sequence and the probability distribution of the next residue derived from similar, yet different, sequences, as summarized in a PSSM. With 975MB required GPU memory to load the final pre-trained model, the average inference time takes around 1.03 second to embed a human protein sequence segment of size 512 residues. The embeddings' prediction power was evaluated on three downstream tasks: secondary structure, subcellular localization, and membrane vs. soluble protein predictions.

Even if I did not succeed in reaching the initial goal of outperforming SeqVec, I did achieve a performance close to SeqVec with half the embedding dimensions (i.e. 3 x 512). Additionally, the results have confirmed a finding by Jozefowicz et al. [37], who did empirically prove that, when trying to fit an LSTM network architecture on very large and complex datasets, the size of the LSTM matters. In general, and from the downstream tasks perspective, there was no evidence that adding PSSM input will improve the knowledge that LMs could learn from a sequence. Different hypotheses can be drawn here: SeqVec was able to learn certain patterns already without PSSMs, the dataset was too small for training an LM, I should have trained the whole architecture from scratch (i.e. without transfer learning), I should have used larger capacity LSTMPs, or I should have simply trained longer.

ACKNOWLEDGMENT

The author would like to thank Michael Heinzinger, Violetta Cavalli-Sforza, and Burkhard Rost for their insights and feedback on this piece of work.